\documentclass[prb,twocolumn,showpacs,floatfix]{revtex4-1}
\usepackage{graphicx}
\usepackage{times}
\usepackage{color}
\input pdfcolor.tex
\usepackage{graphicx}
\usepackage{epsfig,color}
\usepackage{amsfonts,amssymb}
\usepackage{amsmath}
\usepackage{color}
\usepackage{epstopdf}
\definecolor{refcolor}{rgb}{1.0,0.0,0.0}

\newcommand{\be}{\begin{equation}}
\newcommand{\ee}{\end{equation}}   
\newcommand{\bea}{\begin{eqnarray}}
\newcommand{\eea}{\end{eqnarray}}
\newcommand{\ba}{\begin{array}}
\newcommand{\ea}{\end{array}}

\renewcommand{\r}{{\bf r}}

\renewcommand{\k}{{\bf k}}
\newcommand{\Q}{{\bf Q}}

\begin{document}

\title{Roles of anisotropic and unequal gaps in the quasiparticle interference of superconducting iron pnictides}
\date{\today}

\author{Dheeraj Kumar Singh}
\email{dheerajsingh@hri.res.in} 
\affiliation{Harish-Chandra Research Institute,
Chhatnag Road, Jhunsi, Allahabad 211019, India\\
\& Homi Bhabha National Institute, Training School Complex,
Anushakti Nagar, Mumbai 400085, India}
\begin{abstract}
We investigate the role of gap characteristics such as anisotropy and inequality of the gaps
in the quasiparticle interferences of iron pnictides using a five-orbital tight-binding model. 
We examine how the difference in the sensitivities exhibited by 
the sign-changing and -preserving $s$-wave superconductivity in an annular region
around ($\pi, 0$), which can be used to determine the sign change
of the superconducting gap, gets affected when the gaps are unequal on the electron 
and hole pocket. In addition, we also discuss how robust these 
differentiating features are on changing the quasiparticle energy or when the gap is
anisotropic.
\end{abstract}

\maketitle
\newpage
\section{Introduction}
Iron-based superconductors are prototype materials for 
the multiorbital systems exhibiting magnetism and superconductivity \cite{kamihara}. Like cuprates, 
doping either with electrons or holes leads
to the suppression of magnetism and subsequently to the appearance of
superconductivity \cite{dai,avci}. The symmetry of the superconducting (SC) gap function in
this class of materials has been subjected to several theoretical 
\cite{tesanovic,chubukov,singh,beori,mazin,wang,thomale,kuroki,graser,maier} and 
experimental investigations, which is suggested to be extended or sign-changing 
$s$-wave ($s^{+-}$). Experimental determination of 
the sign change is carried out ususally by the phase-sensitive measurements using the 
inelasic-neutron scattering (INS) \cite{lumsden,maier1}, the spectroscopic imaging-scanning
tunneling microscopy (SI-STM) \cite{hanaguri} etc. Latter is a powerful experimental tool and it has
been used to investigate the SC gapstructure \cite{akbari1,akbari2} as well as the normal-state 
bandstructure \cite{lee} in various systems. Recently, it has unraveled the 
existence of highly anisotropic nanostructures in the 
SDW state of the electron-doped iron pnictides \cite{chuang,allan,knolle}. 

Quasiparticle interference (QPI) obtained by the SI-STM depends 
on the sign change of the SC order parameter across the Fermi surfaces (FSs). 
This is because the order parameter enters into the coherence factor representing
the cooper pair through which the quasiparticle scattering takes place. 
In the presence of the non-magnetic impurities, the coherence factor is
vanishingly small for the scattering vectors connecting those part of
the FSs that have same sign of the order 
parameter. Therefore, the QPI pattern is peaked for only those 
momenta that connect the part of the Fermi surfaces with the opposite 
sign of the order parameter \cite{pereg,maltseva,sykora}. However, the situation is reversed 
when the impurities are magnetic. In the iron-based superconductor (IBS), QPI in the 
presence of the magnetic field shows a suppression of the intensity for
the scattering vector connecting the electron and hole FSs separated by
the magnetic ordering wavevector $\Q$ = ($\pi, 0$) \cite{hanaguri}.

According to the experiments, the 
electron FS and the inner hole FS are subject to the 
opening of a lager gap whereas a smaller gap is present at the outer hole FS
\cite{daghero,tortello,ding,evtushinsky}. Moreover, an anisotropic gap has been reported on the 
electron and hole FSs \cite{yoshida,umezawa}. These 
characteristics are expected to play a very important
role in the QPI patterns.

QPIs in the band model \cite{akbari2} as well as in the orbital 
models \cite{zhang} have been investigated with a focus on the difference
of signature of $s^{+-}$- and $s^{++}$-wave superconductivity.
Some of the above mentioned features for the QPIs when the 
impurities are either magnetic or non-magnetic have been
illustrated for the quasiparticle energy 
$\omega \sim$ $\Delta$ (SC gap). Particularly in the orbital model,
the difference in the patterns near ($\pm\pi, 0$) and 
(0, $\pm\pi$) for the $s^{+-}$- and $s^{++}$-wave states 
was shown to exist within a small energy window around 
$\omega \sim |\Delta|$, where the SC gap was chosen
to be $\Delta$ $\sim 100$meV significantly larger 
than the realistic gap size $\Delta \lesssim 20$meV. 
Furthermore, the role of different gaps on the electron 
and the hole pockets as well as the gap anisotropy was also not explored. 
A different group of authors using the same model found
no noticeable difference between the QPI patterns due to 
the $s^{+-}$- and $s^{++}$-wave state, therefore making 
the role of QPI ambiguous in distinguishing the $s^{+-}$- 
from $s^{++}$-wave state \cite{yamakawa}. They examined $Z$ map or the Fourier transform (FT) of 
($g(\r,E)/g(\r,-E)$) instead of the FT of the tunneling 
conductance $g$ map $(g(\r,E))$ a quantity directly 
proportional to the local density of state.

In this paper, we focus on the difference in the $g$ map 
due to the magnetic and non-magnetic impurities instead of 
$Z$ map. We show that the difference can be used to determine 
the sign change of SC order parameter in an unambiguous manner.
We find that (i) the $s^{+-}$- and $s^{++}$-wave states 
exhibit robust pattern of differences within a large part of
the energy region $\omega \le \Delta$, where $\Delta$ is
of the order of observed gap in the experiments. The differences 
drop rapidly with the quasiparticle energy. (ii) The differentiating
features get enhanced at lower energy when either of the gaps on the 
electron and hole pocket is reduced. (iii) On the other hand, these 
features are relatively weaker in the case of anisotropic gap.
The plan of the paper is as follows. In section II, we describe
the procedure to obtain $g$ map. We present results on $g$ maps
in the section III due to the non-magnetic and magnetic impurities for both type 
of $s$-wave superconductivity. Finally, we present our conclusions
in the section IV.
\section{Theory}
\begin{figure}[]
\centerline{
\includegraphics[width=8.2cm,height=10.0cm,angle=0]{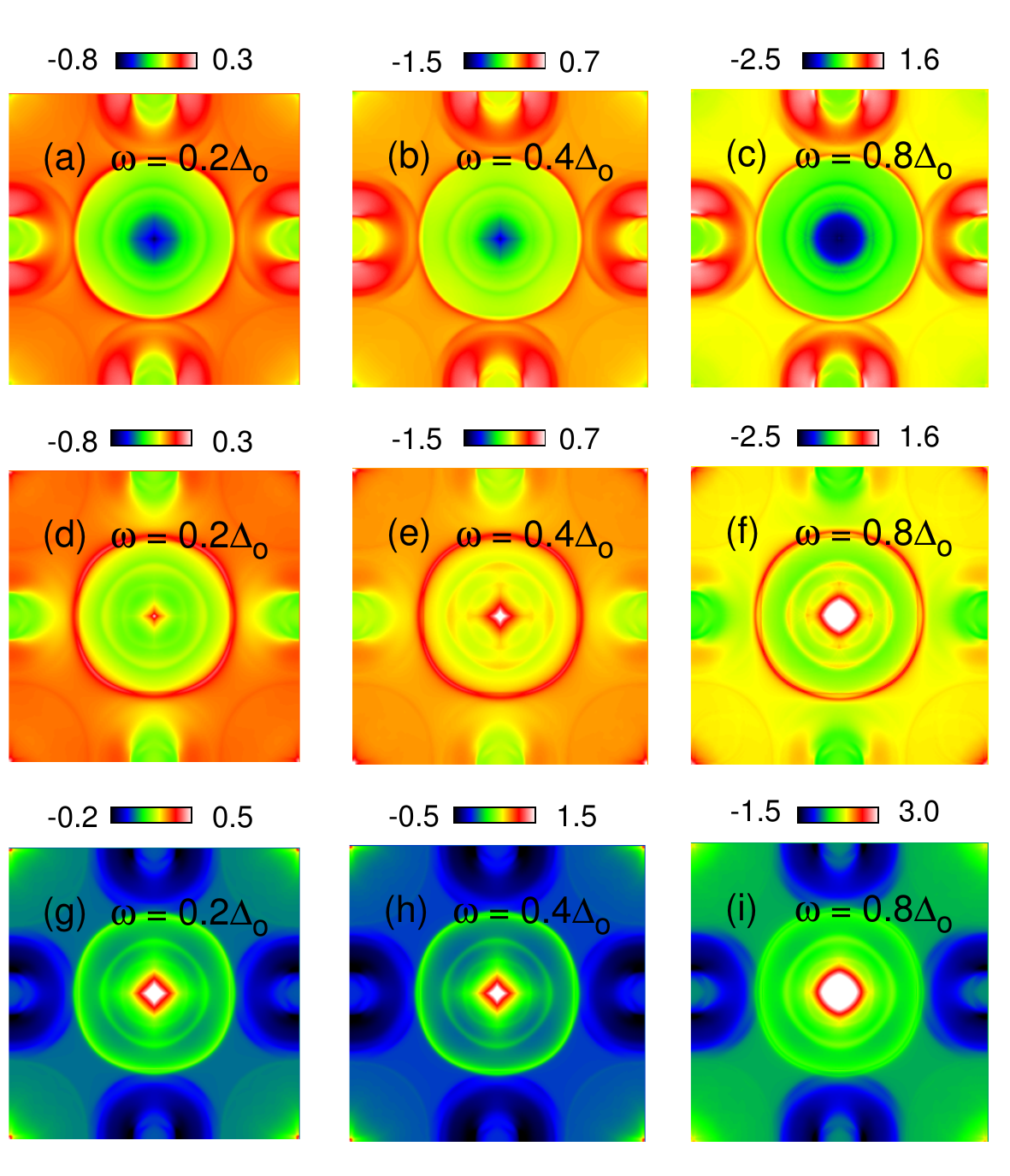}
}
\vspace{0cm}
\caption{QPI patterns for the $s^{+-}$-wave state when the impurity is (a)-(c) non magnetic
and (d)-(e)magnetic. (g)-(i) The difference in the QPI patterns due to the
magnetic and non-magnetic impurities.}
\label{spm1}
\end{figure} 
\begin{figure}[]
\centerline{
\includegraphics[width=8.2cm,height=9.8cm,angle=0]{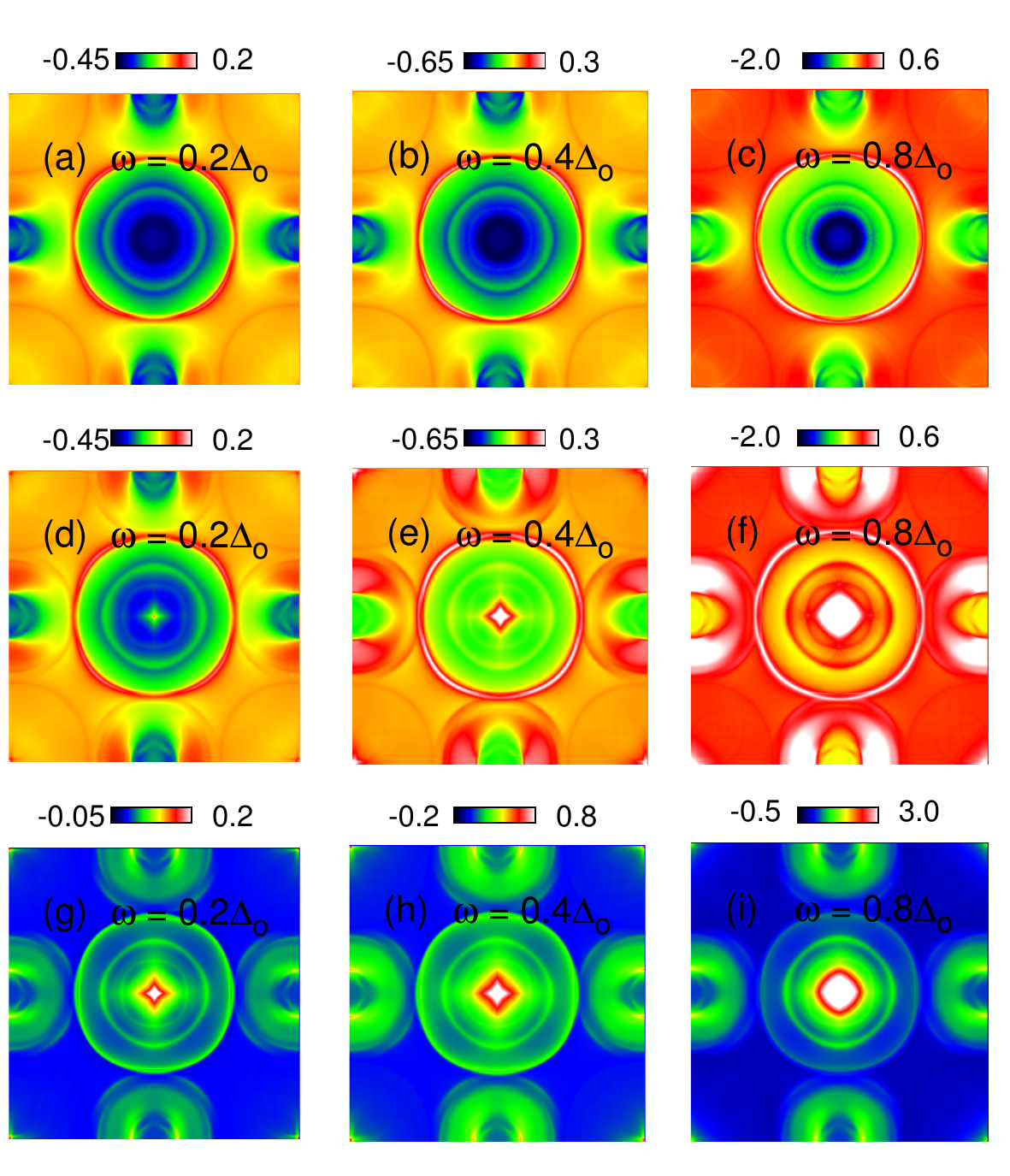}
}
\vspace{0cm}
\caption{Same as in Fig. \ref{spm1} but for the $s^{++}$-wave state.}
\label{spp1}
\end{figure}

In order to investigate QPIs in the superconducting state, we consider
following mean-field Hamiltonian within a five-orbital Hamiltonian
\begin{eqnarray} 
%\begin{split}
 \mathcal{H} &=& \sum_{\k} \Psi^{\dagger}(\k) \hat{H}(\k) \Psi(\k) \nonumber \\
  &=& \sum_{\k} \Psi^{\dagger}(\k)
\begin{pmatrix}
\hat{\varepsilon}({\k}) & \hat{\Delta} \\
\hat{\Delta} & -\hat{\varepsilon}({\k})
\end{pmatrix}
\Psi (\k),
%\end{split}
\end{eqnarray}
where the electron field operator is defined within the Nambu formalism as $\Psi^{\dagger}_{\k \uparrow} = 
(d^{\dagger}_{{\bf k}1\uparrow},d^{\dagger}_{{\bf k}2\uparrow},...d_{-{\bf k}1\downarrow},d_{-{\bf k}2\downarrow}...)$ with 
subscript indices 1, 2, 3, 4, and 5 standing for the orbitals $d_{3z^2-r^2}$, $d_{xz}$, $d_{yz}$, 
$d_{x^2-y^2}$, and $d_{xy}$, respectively. $\hat{\varepsilon}({\k})$ is a 5$\times$5 hopping matrix \cite{ikeda}. 
$\hat{\Delta}(\k)$ is a 5$\times$5 
diagonal matrix, where interorbital 
pairings have been neglected for simplicity. Elements of $\hat{\Delta}(\k)$ are given as 
${\Delta}^{ii}(\k) = {\Delta}^{ii}_o \cos k_x \cos k_y$ and $|{\Delta}^{ii}_o \cos k_x \cos k_y|$ for the $s^{+-}$- and the
$s^{++}$-wave SC states, respectively.

Impurity-induced contribution to the full Green's function is given by 
\be
\delta \hat{G}(\k, \k^{\prime}, \omega) = \hat{G}^0(\k,\omega) \hat{T}(\omega) \hat{G}^0(\k^{\prime},\omega)
\ee
using standard perturbation theory. Here $\hat{G}^0(\k,\omega) = (\hat{\bf I}- \hat{H}(\k))^{-1}$.
$\hat{\bf I}$ is a 10$\times$10 identity matrix. Furthermore,
\be
\hat{T}(\omega) = (\hat{\bf 1} - \hat{V} \hat{\mathcal{G}}(\omega))^{-1}\hat{V},
\ee
with 
\be
\hat{\mathcal{G}}(\omega) = \frac{1}{N} \sum_{\k} \hat{G}^{0}(\k, \omega)
\ee
and
\be
\hat{V} = V_{\circ}
\begin{pmatrix} 
\hat{\bf 1} & \hat{{\rm O}} \\
\hat{{\rm O}} & \pm \hat{\bf 1}  
\end{pmatrix}.
\ee
$\hat{\bf 1}$ and $\hat{{\rm O}}$ are 5$\times$5 identity matrix and null matrices, respectively. +(-) sign in 
front of the identity matrix is for the magnetic (non-magnetic) impurities, respectively. $g$ map or the
fluctuation $\delta N({\bf q},\omega)$ in the LDOS due to a single delta-like impurity scattering is given by 
\be
\delta N({\bf q},\omega) = \frac{i}{2\pi} \sum_{\k} g(\k, {\bf q},\omega)
\ee
with
\be
g(\k, {\bf q},\omega) =  \sum_{i\le5} (\delta \hat{G}^{ii}(\k,{\bf k^{\prime}},\omega)-  \delta \hat{G}^{ii*}({\bf k^{\prime}},\k,\omega)),
\ee
where ${\bf k} - {\bf k}^{\prime} = {\bf q}$.

In the following, the strength of impurity potential $V_{\circ}$ is set to be 200meV. Although, 
we consider a single impurity, the bandfilling $n_e$ is fixed at 6.1. A mesh size
of 300$\times$300 in the momentum space is used for all the calculations. $\Delta_o$ is
taken same for each orbital and it is set to be 20meV throughout. To facilitate a better comparison
between the role of the gap sizes and 
anisotropy, the range of QPI intensity is fixed for each of the $s^{++}$- and $s^{+-}$-wave states.
\section{Results}
\begin{figure}[]
\centerline{
\includegraphics[width=8.2cm,height=10.0cm,angle=0]{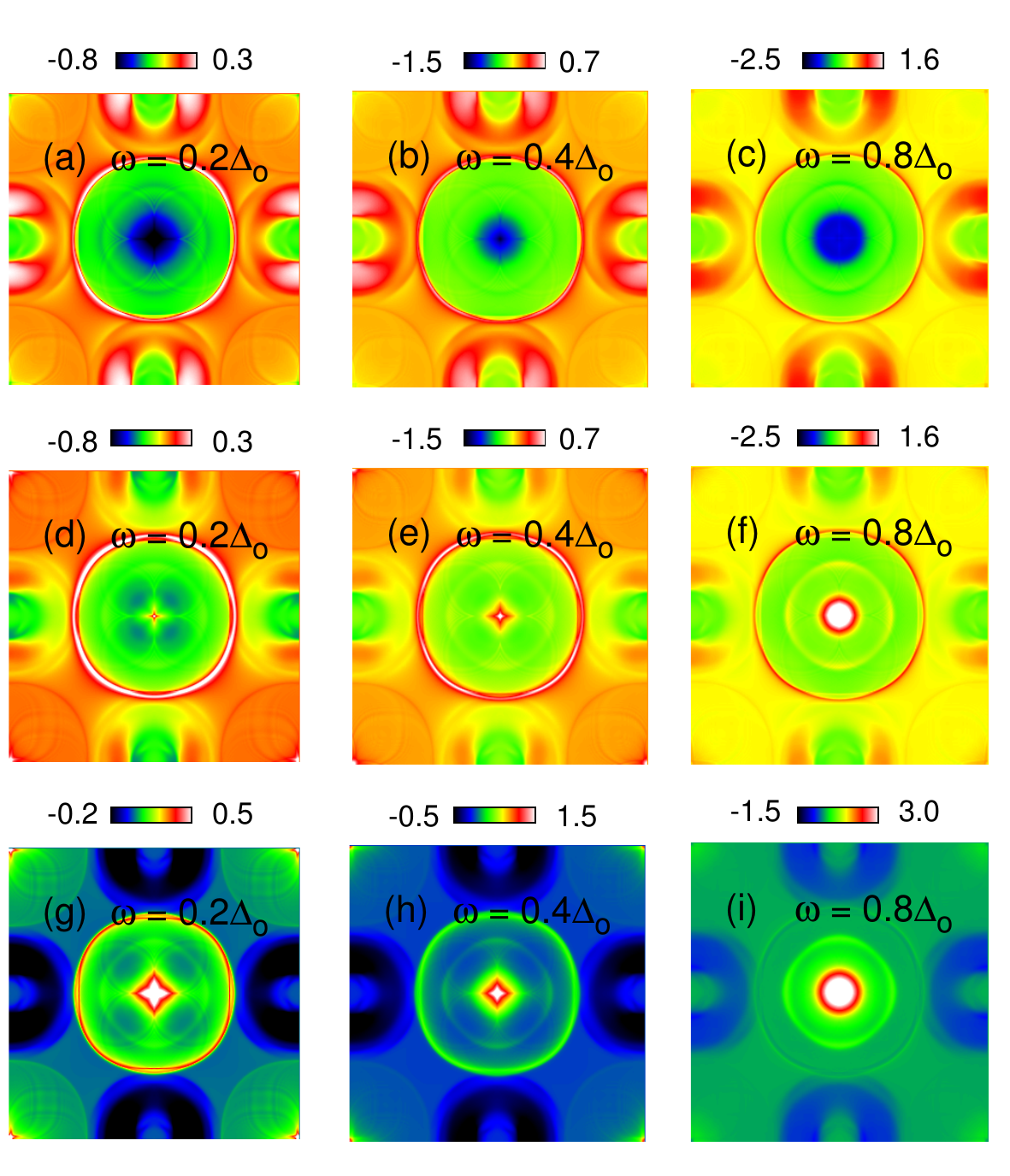}
}
\vspace{0cm}
\caption{QPI patterns for unequal gaps $2\Delta_e = \Delta_h = \Delta_o $ in the 
$s^{+-}$-wave state when the impurity is (a)-(c) non magnetic and (d)-(e)
magnetic. (g)-(i) The difference in the QPI patterns due to the
magnetic and non-magnetic impurities.}
\label{spm2}
\end{figure} 
\begin{figure}[]
\centerline{
\includegraphics[width=8.2cm,height=9.8cm,angle=0]{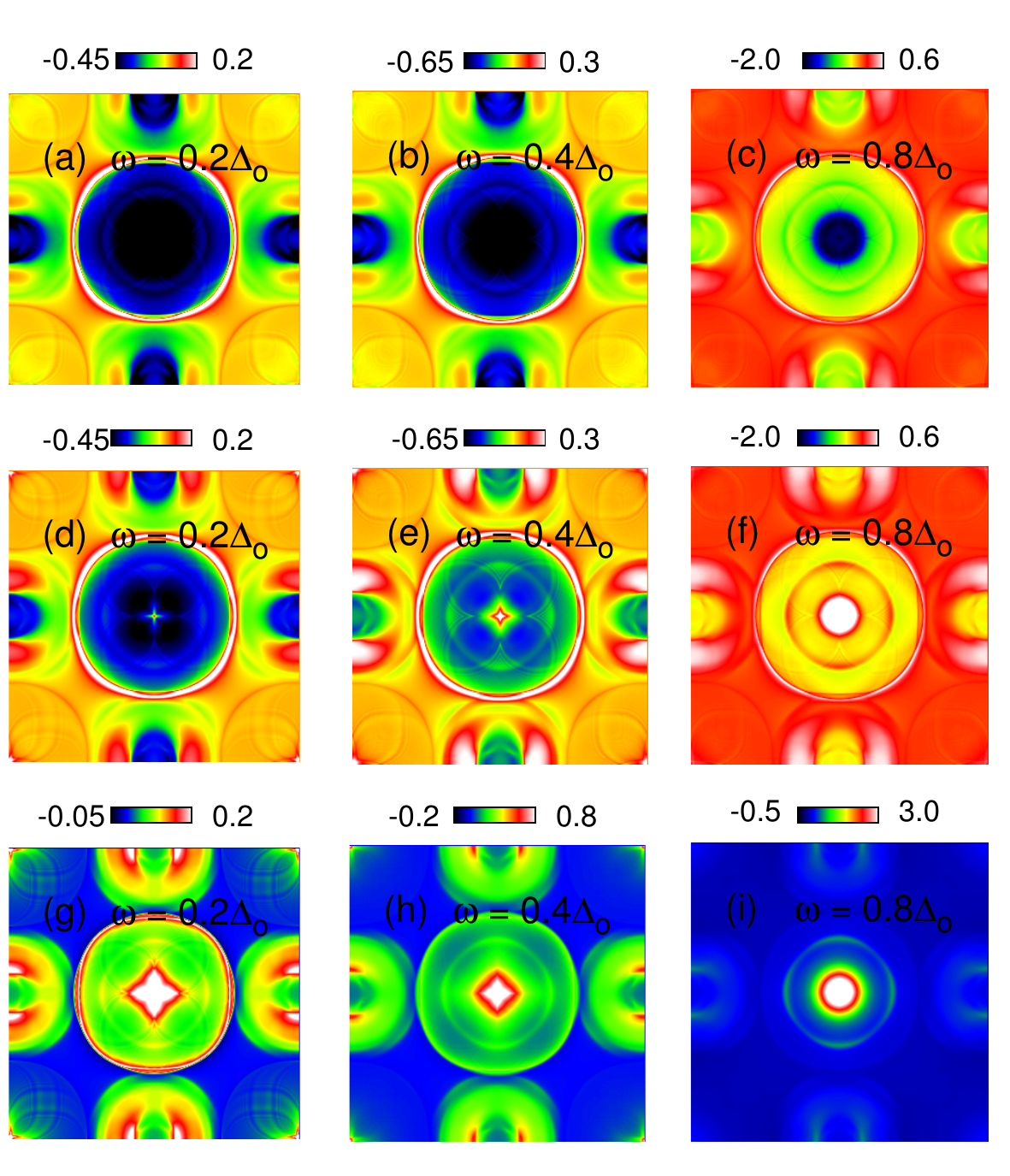}
}
\vspace{0cm}
\caption{Same as in Fig. \ref{spm2} but for the $s^{++}$-wave state.}
\label{spp2}
\end{figure}
Constant energy contours (CCEs) in the $s^{++}$- 
and $s^{+-}$-wave SC states consist of two concentric pockets 
around (0, 0) and another pocket around ($\pi, 0$), which results in several
sets of leading scattering vectors. Most of them can be put into 
three broad catergories. First one consists of
intrapocket and interpocket scattering vectors arising due 
to the pockets around (0, 0). Second set involves the scattering 
vectors connecting the pockets around (0, 0) and ($\pi, 0$). Third one is the set of scattering vectors 
connecting the CCEs around ($0, -\pi$) and ($\pi$, 0). This is a non exhaustive list. For instance, 
the set of vectors connecting the pocket around ($\pi, \pi$) to other
pockets will also contribute to the scattering processes.

First of all, we consider 
equal gap $\Delta_e = \Delta_h = \Delta_o$, where $\Delta_e$ and $\Delta_h$ are the SC 
gaps along the electron and hole pockets, respectively. Fig. \ref{spm1}(a)-(c) and \ref{spm1}(d)-(f) show the QPI patterns in the $s^{+-}$-wave SC state for the 
non-magnetic and magnetic impurities, respectively. 
Results are obtained for several energy values $\omega = 0.2\Delta_o, 
0.4\Delta_o, {\rm and}\, 0.8\Delta_o$. Two important differences
in the QPI patterns due to the non-magnetic and magnetic 
impurities can be noticed. First, an annular region around ($\pi$, 0) and other symmetrically 
equivalent points are comparatively more intense when the impurity is non-magnetic. Secondly, the sign of the 
peak at (0, 0) in the case of non-magnetic impurity is opposite to that due to the magnetic impurity. 
More clarity can be found by plotting the differences of QPI patterns due to the magnetic and nonmagnetic impurities
as shown in the Fig. \ref{spm1}(g)-(i), where the peaks are positive and negative around 
(0, 0) and ($\pi, 0$), respectively. It is important to note that the differences in the features are
more pronounced in the vicinity of $\omega \sim \Delta_o$ and although they are also present at 
lower energy but weakened relatively.

Fig. 2 shows the QPI patterns when the SC gap has $s^{++}$-wave symmetry. As can be seen from Fig. \ref{spp1}(a)-(c),
the largest negative peak in the non-magnetic case occurs near (0, 0) a feature also present in the
patterns for the $s^{+-}$-wave state. However, the intense and positive annular peak structure around ($\pi, 0$) occurs 
now when the impurity is magnetic (Fig. \ref{spp1}(d)-(f)). 
This is seen more vividly in the differences of the QPI patterns (Fig. \ref{spp1} (g)-(i)), which is positive 
in the annular region around ($\pi, 0$). Therefore, the 
differences in the QPI patterns due to the magnetic and non-magnetic impurities can be 
used to determine the sign change of the SC gap.

\begin{figure}[]
\centerline{
\includegraphics[width=8.2cm,height=10.0cm,angle=0]{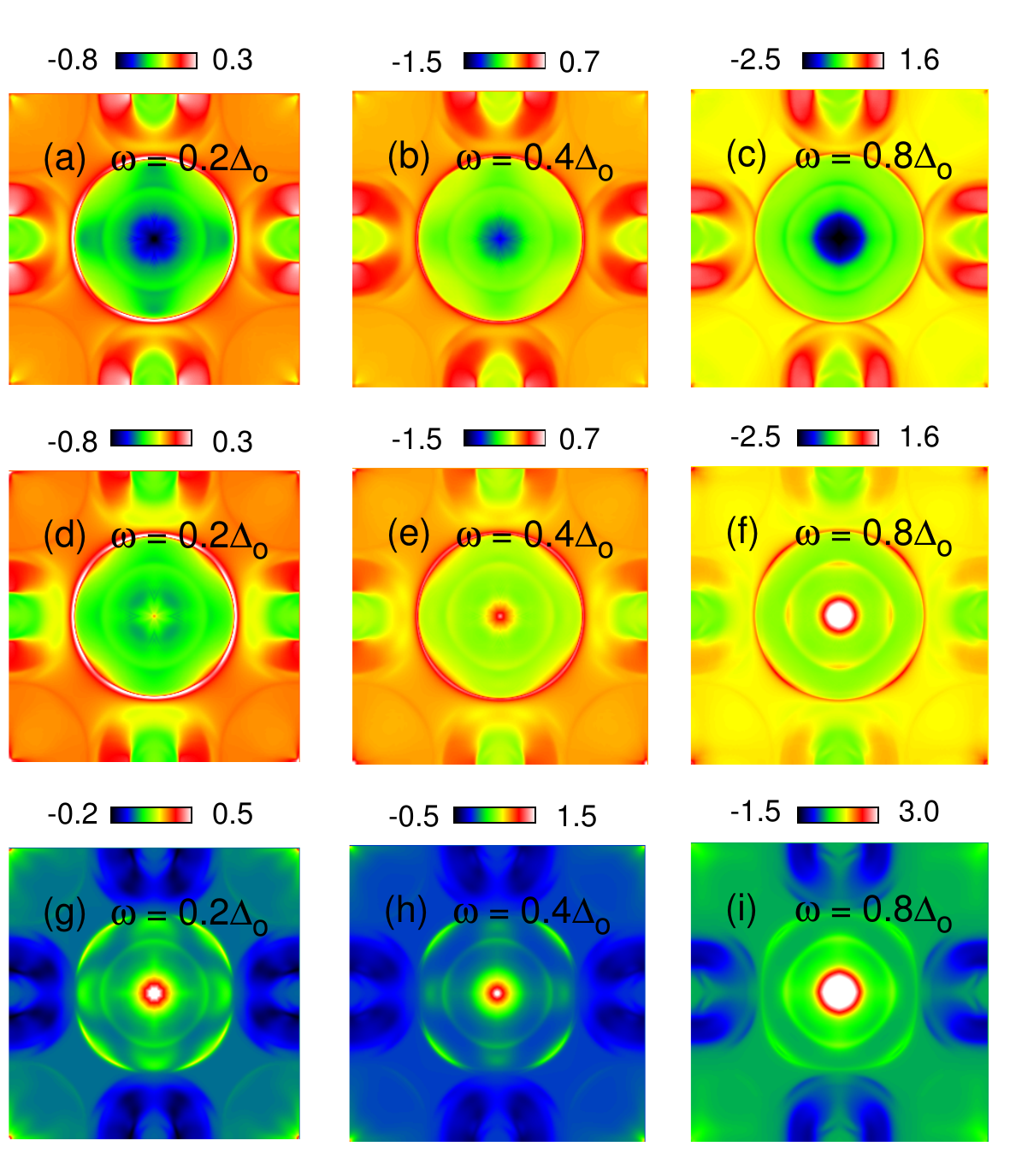}
}
\vspace{0cm}
\caption{QPI patterns for the anisotropic gap $\Delta_e = \Delta_o(1+\cos2\theta)$ along the electron pocket in the $s^{+-}$-wave state when the impurity is non magnetic
(a)-(c) and magnetic (d)-(e). (g)-(i) The difference in the QPI patterns due to the
magnetic and non-magnetic impurities.}
\label{spm3}
\end{figure} 
\begin{figure}[]
\centerline{
\includegraphics[width=8.2cm,height=9.8cm,angle=0]{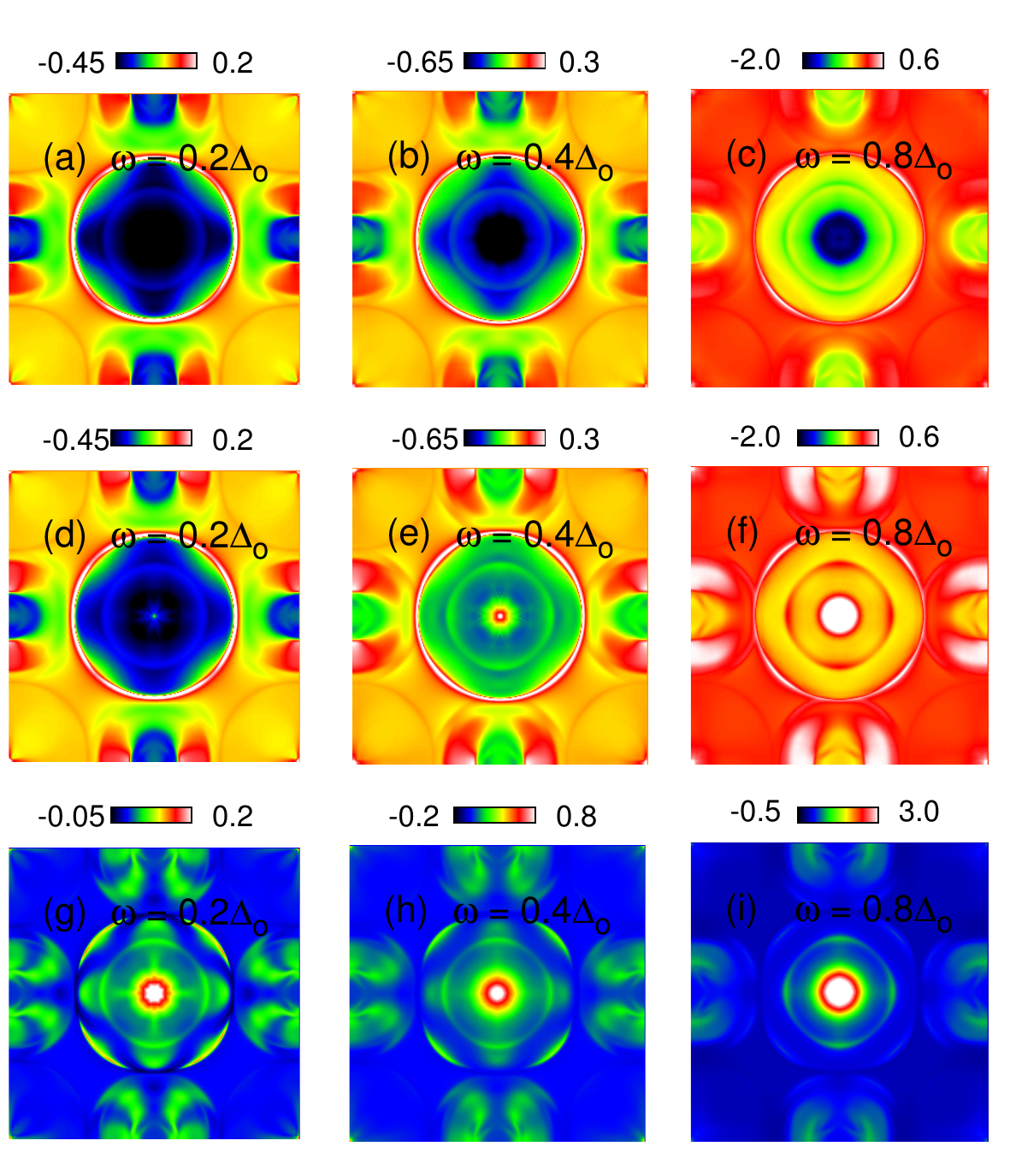}
}
\vspace{0cm}
\caption{Same as in Fig. \ref{spm3} but for the $s^{++}$-wave state.}
\label{spp3}
\end{figure}
For simplicity and to focus on individual role, we examine the unequal gaps and anisotropy separately. 
We consider $2\Delta_e = \Delta_h = \Delta_o$. Similar results are 
obtained for $2\Delta_h =  \Delta_e = \Delta_o$. Fig. \ref{spm2}(a)-(c) show the QPI patterns 
for the unequal gaps when the impurity is non-magnetic. An important 
consequence of reducing the gap along the electron pocket is the suppression of intensity in the annular region
when $\omega \sim \Delta_o$. However, the intensity grows on decreasing $\omega$ in comparison
with the equal gap case. As mentioned earlier, when the gaps are equal,
QPIs are most intense for $\omega \sim \Delta_o$. Therefore, 
when one of the gap is reduced, the energy corresponding to the maximum intensity shifts downwards. As a result the intensity increases at lower energy.
This happens for the annular region around ($\pi, 0$) even if the impurity
is magnetic irrespective of the $s^{++}$- or $s^{+-}$-wave states (Fig. \ref{spm2}(d)-(f), 
Fig. \ref{spp2}(a)-(c), and Fig. \ref{spp2}(d)-(f)). In other words, there is no qualitative difference between the 
QPI patterns due to the magnetic and non-magnetic impurities. However, increase in the 
intensity is more in the case of the non-magnetic and magnetic impurities 
for the $s^{+-}$- and $s^{++}$-wave states, respectively. Consequently in comparison 
with the equal gaps case, differences in 
the patterns for magnetic and non-magnetic impurities are enhanced especially 
at lower energy in both type of SC states (Fig. \ref{spm2}(g)-(i), and Fig. \ref{spp2}(g)-(i)).

Fig. \ref{spm3} and \ref{spp3} show the QPI patterns for the $s^{+-}$- and the $s^{++}$-wave state when the 
SC gap on the electron pocket is anisotropic. The gap is chosen as $\Delta_e = \Delta_o(1+\cos2\theta)$, where
$\theta$ is defined in such a way that $\Delta_e$ is maximum for the point along the line joining 
($\pi, 0$) to ($0, 0$). The most important effect of the anisotropic gap is 
that the difference in the QPI patterns due to the magnetic and the non-magnetic impurities diminishes especially at low energy
in the annular region around ($\pi, 0$) (Fig. \ref{spm3}(a)-(f)). In addition, the peak structures are also modified 
particularly in the annular region. That is well reflected in the plot of
differences (Fig. \ref{spm3}(g)-(i)). Similar results are also obtained in the case of
$s^{++}$-wave state (Fig. \ref{spp3}(a)-(i)). Therefore, the difference 
in the QPI patterns corresponding to the $s^{+-}$- and $s^{++}$-wave state 
also decreases, but the two SC states are still robustly 
distinguishable.

\section{Conclusions} 
In conclusions, we have investigated the QPIs in a five-orbital model of iron pnictides with a focus 
on the roles of gap features such as inequality and anisotropy. We find that the differentiating 
features of the $s^{+-}$- and $s^{++}$-wave state, which mainly consist of the QPI patterns 
in an annular region around ($\pi, 0$), decrease rapidly with
the quasiparticle energy, though they are present even at lower energy. The rapid drop is slowed down 
when one of the gap is reduced, which also results in the widening of the quasiparticle 
energy range wherein the difference in QPI patterns for two types of superconducitivity is enhanced. We also find 
that appearance of the patterns due to magnetic and non-magnetic impurities are qualitatively similar. Moreover, 
it is only the difference in the patterns, which is helpful in ascertaining the sign-change of the gap. On the contrary, 
anisotropic gap leads to the reduction in the differentiating signatures of sign change especially at low energy. 
Although these characteristics are not responsible for introducing any major qualitative change in the QPI patterns, but 
they do affect the QPI peak strength in a significant manner. 


\begin{thebibliography}{0}

\bibitem{kamihara} 
Kamihara Y., J. Am. Chem. Soc., {\bf 130} (2008) 3296.

\bibitem{dai}
Dai P., Hu J. and Dagotto E., {\it Nat. Phys.}, {\bf 8} (2012) 709. 

\bibitem{avci}
Avci S. {\it et. al.}, {\it Phys. Rev. B}, {\bf 85} (2012) 184507.

\bibitem{tesanovic}
Cvetkovic V. and Tesanovic Z., {\it Europhys. Lett.}, {\bf 85} (2009) 37002.

\bibitem{chubukov}
Chubukov A. V., Efremov D. V. and Eremin I., {\it Phys. Rev. B}, {\bf 78} (2008) 134512.

\bibitem{singh}
Singh D. J. and Du M.-H., {\it Phys. Rev. Lett.} {\bf 100}, (2008) 237003. 

\bibitem{beori}
Boeri L., Dolgov O. V. and Golubov A. A., {\it Phys. Rev. Lett.}, {\bf 101}, (2008) 026403.

͑\bibitem{mazin}
Mazin I. I., Singh D. J., Johannes M. D. and Du M. H., {\it Phys. Rev. Lett.}, {\bf 101}, (2008) 057003. 

\bibitem{wang}
Wang F., Zhai H. and Lee D.-H., {\it Phys. Rev. B}, {\bf 81} (2010) 184512. 

\bibitem{thomale}
Thomale R. {\it et. al.}, {\it Phys. Rev. B}, {\bf 80} (2009) 180505.

\bibitem{kuroki}
Kuroki K. \textit{et. al.}, {\it Phys. Rev. Lett.}, {\bf 101} (2008) 87004.

\bibitem{graser}
Graser S., Maier T. A., Hirschfeld P. J. and Scalapino D. J., {\it New J. Phys.}, {\bf 11} (2009) 025016.

\bibitem{maier}
Maier T. A., Graser S., Scalapino D. J., Hirschfeld P. J., {\it Phys. Rev. B}, {\bf 79} (2009) 224510.

\bibitem{lumsden}
Lumsden M. D., {\it Phys. Rev. Lett.} {\bf 102} (2009) 107005. 

\bibitem{maier1}
Maier T. A., Graser S., Scalapino D. J. and Hirschfeld P., 
{\it Phys. Rev. B} {\bf 79} (2009) 134520.

\bibitem{hanaguri}
Hanaguri T., Niitaka S., Kuroki  K. and Takagi H., {\it Science} {\bf 328} (2010) 474.  

\bibitem{akbari1}
Akbari A. and Thalmeier P., {\it Phys. Rev. B} {\bf 88} (2013) 134519.

\bibitem{akbari2}
Akbari A., Thalmeier P. and Eremin I., {\it Phys. Rev. B} {\bf 84} (2011) 134505.

\bibitem{lee}
Lee W.-C., Wu C., Arovas D. P. and Zhang S.-C., {\it Phys. Rev. B}, {\bf 80} (2009) 245439.


\bibitem{chuang}
Chuang T.-M. {\it et. al.}, {\it Science} {\bf 327}, (2010) 181. 

\bibitem{allan}
Allan M. P. {\it et. al.}, Nat. Phys. {\bf 9}, 220 (2013).

\bibitem{knolle}
Knolle J., Eremin I., Akbari  A. and Moessner R., {\it Phys. Rev. Lett.} {\bf 104}, 257001 (2010).

\bibitem{pereg}
Pereg-Barnea T. {\it et. al.}, {\it Phys. Rev. B}, {\bf 78} (2008) 020509.

\bibitem{maltseva}
Maltseva M. and Coleman P., {\it Phys. Rev. B}, {\bf 80} (2009) 144514.

\bibitem{sykora}
Sykora S. and Coleman P., {\it Phys. Rev. B}, {\bf 84} (2011) 054501. 

\bibitem{daghero}
Daghero D. \textit{et. al.} {\it Phys. Rev. B}, {\bf 80}, (2009) 060502(R).

\bibitem{tortello}
Tortello M. \textit{et. al.} {\it Phys. Rev. Lett.}, {\bf 105}, (2010) 237002.

\bibitem{ding}
Ding H. \textit{et. al.}, {\it Europhys. Lett.}, {\bf 83}, 47001 (2008).

\bibitem{evtushinsky}
Evtushinsky D. V. \textit{et. al.}, {\it Phys. Rev. B} {\bf 79},  (2009) 054517.

\bibitem{yoshida}
Yoshida T. {\it et. al.}, {\it Sci. Rep.}, {\bf 4} (2014) 7292. 

\bibitem{umezawa}
Umezawa K. {\it et. al.}, {\it Phys. Rev. Lett.}, {\bf 108}, (2012) 037002.

\bibitem{akbari2}
Akbari A., Knolle J., Eremin I. and Moessner R., {\it Phys. Rev. B}, {\bf 82} (2010) 224506. 

\bibitem{zhang} 
Zhang Y.-Y. {\it et. al.}, {\it Phys. Rev. B} {\bf 80}, (2009) 094528.

\bibitem{yamakawa}
Yamakawa Y. and Kontani H., {\it Phys. Rev. B}, {\bf 92} (2015) 045124. 


\bibitem{ikeda}
Ikeda H., Arita R. and Kunes  J., {\it Phys. Rev. B}, {\bf 81} (2010) 054502.

\end{thebibliography}
\end{document}